\documentclass[reprint,amsmath,amssymb,aps,prx]{revtex4-1}
\usepackage[dvips]{graphicx}
\usepackage{dcolumn}
\usepackage{bm}
\usepackage{hyperref}
\usepackage{color}
\usepackage{cancel}
\usepackage{bmpsize}
\usepackage{amsmath}
\usepackage{amssymb}
\usepackage{booktabs}
\usepackage{nicefrac}

\newcommand{\expect}[1]{\left\langle #1\right\rangle}

\begin{document}

\title{Quantum-inspired annealers as Boltzmann generators\\for machine learning and statistical physics.}

\author{Alexander E. Ulanov$^{1,2}$, Egor S. Tiunov$^{1,2}$ and A. I. Lvovsky$^{1,3,4}$}

\affiliation{$^1$Russian Quantum Center, Skolkovo, Moscow 143025, Russia}
\affiliation{$^2$Moscow Institute of Physics and Technology, 141700 Dolgoprudny, Russia}
\affiliation{$^3$Department of Physics, University of Oxford, Oxford OX1 3PU, UK}
\affiliation{$^4$P. N. Lebedev Physics Institute, Leninskiy prospect 53, Moscow 119991, Russia}

\email{a.ulanov@rqc.ru}
\email{alex.lvovsky@physics.ox.ac.uk}
\date{December 18, 2019}
%
\begin{abstract}

Quantum simulators and processors are rapidly improving nowadays, but they are still not able to solve complex and multidimensional tasks of practical value. However, certain numerical algorithms inspired by the physics of real quantum devices prove to be efficient in application to specific problems, related, for example, to combinatorial optimization. Here we implement a numerical annealer based on  simulating the coherent Ising machine as a tool to sample from a high-dimensional Boltzmann probability distribution with the energy functional defined by the classical Ising Hamiltonian. Samples provided by such a generator are then utilized for the partition function estimation of this distribution and for the training of a general Boltzmann machine. Our study  opens up a door to practical application of numerical quantum-inspired annealers.

\end{abstract}

\maketitle
\vspace{10 mm}

\section{Introduction}
\label{intro}

The Ising model is a cornerstone of various fields of science  ranging from magnetism \cite{Baxter2014} and quantum field theory \cite{Zuber1977} to combinatorial optimization \cite{McMahon2016} and finance \cite{Sornette2014}. This model analyzes magnetic interactions in a set of spin-$\frac12$ particles. The energy of each spin configuration with external fields is given by 
\begin{equation}\label{energyBM}
E(\textbf{s}) = -\frac{1}{2}\textbf{s}^\mathrm{T}\hat J\textbf{s}-\textbf{b}^\mathrm{T}\textbf{s},
\end{equation}
where $ \textbf{s} \in \left\lbrace -1,1\right\rbrace ^{N} $ is the vector of spins values, $\hat J$ the coupling matrix,  $\textbf{b}$ the bias vector and $N$ is the number of interacting spins. 
The Ising problem consists in finding the ground state or low-energy spin configurations of the energy functional \eqref{energyBM}. This is known to be an NP-hard combinatorial problem \cite{Barahone1982} and multiple classical numerical algorithms \cite{Kirkpatrick1983, Bilbro1989, Benlic2013}, neural-network based methods \cite{Smith1999, Barrett2019}, quantum hardware devices \cite{McMahon2016,Harris2018,Inagaki2016}  and quantum-inspired numerical algorithms \cite{King2018, Kalinin2018, Tiunov2019} have been developed to approximately solve it.

The joint probability corresponding to each spin configuration is described by the Boltzmann distribution
\begin{equation}\label{probability}
p(\textbf{s}) = \frac{e^{-\beta E(\textbf{s})}}{Z},
\end{equation}
where $\beta = 1/T$ is the inverse temperature of the system and the normalizing constant $Z = \sum_{\textbf{s}} {e^{-\beta E(\textbf{s})}}$  is referred to as the partition function. Knowledge of the partition function is important in multiple tasks ranging from statistical physics \cite{Wu2019} to machine learning \cite{Ma2013}. While the unnormalized probability $e^{-\beta E(\textbf{s})}$ is easy to compute for any spin configutation, the exact calculation of the partition function is intractable for a large and fully connected spin system. Therefore, approximate methods are usually employed. One of the most popular approaches is the iterative solution of a set of mean-field equations, such as the naive mean-field approach, Bethe \cite{Bethe1935} and Thouless-Anderson-Palmer \cite{Thouless1977} approximations. While relatively easy to implement, mean-field-based methods \cite{Jordan1999}  usually perform poorly when dealing with systems having strong, long-range correlations between spins. 

An alternative approach is by generating unbiased samples from the true Boltzmann distribution, which can then be used to estimate the partition function either by direct summation or using more advanced algorithms such as annealed importance sampling (AIS) \cite{Neal1998}. Boltzmann sampling is an important problem in its own right, relevant to multiple tasks in machine learning \cite{Hinton2006}, many-body physics \cite{Carleo2017}, quantum-state tomography \cite{Torlai2018, TiunovQST2019}, chemistry \cite{Zhu2002} and protein folding \cite{Rizzuti2013}. 

Software-based sampling algorithms include Markov chain Monte Carlo (MCMC) \cite{Newman1999} or simulated annealing \cite{Kirkpatrick1984}; however, these methods are prone to get trapped in local optima. Recently, this family was joined by machine-learning (ML) based algorithms, namely Variational Autoregressive Networks (VAN) \cite{Wu2019} and Boltzmann Generator Networks (BGN) \cite{Noe2019}, which have been demonstrated to be  efficient for both approximate Boltzmann sampling and partition function estimation.

In 2014, Dumoulin \textit{et al.} \cite{Dumoulin2014} proposed to use quantum hardware --- specifically, the quantum annealer --- as a fast source of samples from a given Boltzmann distribution. This study analyzed the main limitations imposed by existing quantum hardware (such as D-Wave) among which are noise in parameters, limited parameter range, and restrictions in available architectures. The proposed method was experimentally implemented using D-Wave 2X quantum annealer \cite{Benedetti2016,Benedetti2017} for the training of a Boltzmann machine \cite{Hinton2006,Teh2006}. However, the above mentioned limitations in existing quantum annealers limited the study to low-dimensional datasets. 

A novel device that appears promising in this context is the  optical coherent Ising machine (CIM), which was shown  \cite{McMahon2016, Inagaki2016} to find good ground state approximations for classical, fully connected Ising systems of sizes up to  2000 spins. In a separate study, CIM was shown to significantly outperform the D-Wave annealer in application to densely connected Ising systems \cite{CIMvsDwave}. This technology inspired a range of classical algorithms, some of which ---  the noisy mean-field annealing (NMFA) \cite{King2018} and the CIM simulation (SimCIM) \cite{Tiunov2019} --- were shown to outperform the CIM in terms of both the computational speed and mean sample energy. 
A further advantage of these algorithms is the ability to operate with Ising Hamiltonian with the coupling and bias matrix elements described by arbitrary real numbers. 
However, to the best of our knowledge, neither CIM nor any of the quantum-inspired algorithms (NMFA or SimCIM) have been applied to Boltzmann sampling yet.

An important application of the Boltzmann sampling and the partition function estimation is the setting known as the inverse Ising problem \cite{Nguyen2017} in statistical physics or Boltzmann machine (BM) \cite{Hinton1983, Hinton2002} in ML. It consists in finding the coupling matrix $\hat{J}$ and bias vector $\textbf{b}$ entering the Boltzmann probability law \eqref{probability}, which maximizes the likelihood of a specific set of spin configuration samples. The samples are utilized  to estimate the intractable terms in the gradients arising in the training.

In this paper, we find that the quantum-inspired numerical annealer SimCIM \cite{Tiunov2019} is capable of drawing high-quality unbiased samples from a Boltzmann probability distribution associated with the Ising model. We demonstrate the training of a fully-visible and fully-connected BM using samples provided by SimCIM on two training sets. The first dataset (Bars\&Stripes, BAS) is of low dimension, which permits exhaustive search over all possible spin configuration, thereby enabling direct quality estimation and comparison of various sampling methods. Second, we train the BM on the high-dimensional dataset of handwritten digit images MNIST. The corresponding Ising graph is fully connected and has a dimension of 794. In addition to character recognition, we numerically estimate the partition function $Z$ of this system. In the latter task, we use AIS with SimCIM as the sampler for all intermediate distributions.

SimCIM implementation for Boltzmann sampling is beneficial compared to VAN or BGN because, in order to draw samples, SimCIM requires no knowledge about the system of interest apart from the spin interaction strength. Unlike VAN or BGN, SimCIM does not require specific neural network training for each update of the coupling matrix and bias vector and each run of the partition function estimation, which could be very demanding for high dimensional systems. On the other hand, in comparison with analog annealers such as D-Wave or CIM, SimCIM is advantageous in that it supports real-valued and fully-connected high-dimensional coupling matrices and could be executed on a classical computer.


\section{Review of previous results and methods} \label{Recapsec}
In this section, we give a brief recap of previous results and techniques that are relevant to our work.
\subsection{General Boltzmann Machines} \label{BMsec}
The Boltzmann machine (BM) is a stochastic energy-based data model which was introduced at the dawn of ML by Hinton and Sejnowski \cite{Hinton1983}. BM is the primary component of deep belief networks \cite{Hinton2006} and deep Boltzmann machines \cite{Salakhutdinov2009}. 
Mathematically, the BM can be represented as a complete graph whose nodes represent binary units that can take on the values of $\pm 1$. The nodes are connected by edges, each of which is associated with an arbitrary real value. The graph is assigned the ``energy" described by Eq.~\eqref{energyBM}. 

To apply the BM for machine learning --- for example, character recognition --- we associate every pixel of the bitmap containing the character with a node of the BM. The BM parameters (coupling matrix and biases) are ``trained", i.e. assigned such values that the energy \eqref{energyBM} associated with the bitmaps within the training set is significantly lower than that for an arbitrary bitmap. Then, when posed with a task of recognizing whether a particular unknown bitmap resembles those from the training set, the energy \eqref{energyBM} for that bitmap is calculated. A low value would indicate a high likelihood of the affirmative answer. 

In order to facilitate the training, we further associate the energy with the Boltzmann-like probability given by Eq.~\eqref{probability} with $\beta=1$. The optimization objective during the BM training is the maximization of the total log-likelihood $ \mathcal L=\sum_{\textbf{S}}\log p(\textbf{s})$ of all bitmaps of the training set $\textbf{S}$. We then have \cite{HintonPracticalGuide}
\begin{align}
\nonumber
\frac{\partial \mathcal L}{\partial J_{ij}}&=\sum_{\textbf{S}}\left(s_i s_j-\langle s_i s_j\rangle_{\{\textbf{s}\}}\right)\\
\frac{\partial \mathcal L}{\partial b_i}&=\sum_{\textbf{S}}\left(s_i-\langle s_i\rangle_{\{\textbf{s}\}}\right),\label{trainderiv}
\end{align}
which gives rise to a conceptually straightforward gradient descent update rule
\begin{align}
\Delta\hat J&=\xi\left(\langle \bold{s}\bold{s}^\mathrm{T}\rangle_{\textbf{S}}-\langle \bold{s}\bold{s}^\mathrm{T} \rangle_{\{\textbf{s}\}}\right);\nonumber \\
	\Delta b&=\xi\left(\langle \bold{s}\rangle_{\textbf{S}}-\langle \bold{s} \rangle_{\{\textbf{s}\}}\right),\label{update}
\end{align}
where $\xi$ is the learning rate. In Eq.~\eqref{update}, the first term is the expectation with respect to the training set and often called the positive phase of the update, while the second term is the expectation over the model's probability distribution  $p(\bold{s})$ and called the negative phase. 

Practical implementation of this gradient descent is complicated by the fact that 
calculating the negative phase requires exhaustive search over all possible configurations of  $\textbf{s}$, and is hence intractable for a classical computer. Therefore it is usually replaced with a numerical estimation using a set of configuration samples drawn from the distribution \eqref{probability}. In this case, Eqs.~(\ref{update}) become
\begin{align}\label{approx_updates}
\Delta \hat J &\approx \xi \left (\frac{1}{N}\sum_{n=1}^{N}\bold{s}_n\bold{s}_n^\mathrm{T} - \frac{1}{M}\sum_{m=1}^{M}\hat{\bold{s}}_m\hat{\bold{s}}_m^\mathrm{T}\right);\nonumber \\
\Delta b &\approx \xi \left (\frac{1}{N}\sum_{n=1}^{N}\bold{s}_n - \frac{1}{M}\sum_{m=1}^{M}\hat{\bold{s}}_m\right),
\end{align}
where $N$ is the number of elements in the training set, $M$ the number of drawn samples.  

Traditionally, in the context of BM training, the sampling has been approached using MCMC \cite{Salakhutdinov2009}. This approach is however too slow to be practical because sometimes it requires an extremely large number of steps to give unbiased equilibrium samples. This motivates the interest to quantum annealers and their simulators in the context of BM training. 
\subsection{Annealed importance sampling}
\label{AISsec} 

It is not required to know the exact value of partition function $Z$ for the approximate BM training procedure introduced above. However, this  knowledge of $Z$ is necessary for the evaluation of the training set likelihood $\mathcal L$, and hence permits the estimation of the training quality. While it appears straightforward that the partition function can be evaluated directly by performing summation over a large number of samples, this approach is less efficient than the procedure known as annealed importance sampling \cite{Neal1998}.

Consider two probability distributions $\mathfrak{A}$ and $\mathfrak{B}$ defined as $p_\mathfrak{A}(\mathbf{s}) = p^*_\mathfrak{A}(\mathbf{s})/Z_\mathfrak{A}$ and $p_\mathfrak{B}(\mathbf{s}) = p^*_\mathfrak{B}(\mathbf{s})/Z_\mathfrak{B}$, where $p^* = e^{-E(\mathbf{s})}$ is the corresponding unnormalized probability which is easily computable for both distributions. Assume that $\mathfrak{B}$ is uniform while $\mathfrak{A}$ is the distribution of our interest. Then we can estimate $Z_\mathfrak{A}$ with samples from the uniform distribution using importance sampling
\begin{align} \label{IS}
\frac{Z_\mathfrak{A}}{Z_\mathfrak{B}} &= \sum_{\textbf{s}}\frac{p^*_\mathfrak{A}(\mathbf{s})}{Z_\mathfrak{B}} = \sum_{\textbf{s}}\frac{p^*_\mathfrak{A}(\mathbf{s})}{p^*_\mathfrak{B}(\mathbf{s})}p_\mathfrak{B}(\mathbf{s}) = \expect{\frac{p^*_\mathfrak{A}(\mathbf{s})}{p^*_\mathfrak{B}(\mathbf{s})}}_{\mathbf{s}\sim p_\mathfrak{B}(\mathbf{s})}\nonumber \\
&\approx \frac{1}{M}\sum_{i}^{M}\frac{p^*_\mathfrak{A}(\mathbf{s}_i)}{p^*_\mathfrak{B}(\mathbf{s}_i)} = \frac{1}{M}\sum_{i}^{M}w_i,
\end{align}
where $\mathbf{s}_i$ denotes the samples drawn from the uniform distribution $p_\mathfrak{B}$. This method, while being relatively simple to implement, gives estimates of $Z_\mathfrak{A}$ with high variance especially when the target distribution $p_\mathfrak{A}$ differs significantly from uniform. To reduce the variance of this estimate, a sequence of intermediate distributions $p_{0::N}$, where $p_0 = p_\mathfrak{B}$ and $p_N = p_\mathfrak{A}$ is chosen to gradually transition from the uniform to target distribution. For each pair of consecutive  distributions, the importance weight $ Z_{n+1}/Z_n$ is calculated using Eq. \eqref{IS} and then the partition function of the target distribution is estimated by multiplying
\begin{equation} \label{AIS}
Z_\mathfrak{A} = Z_\mathfrak{B}\prod_{n=0}^{N-1}\frac{Z_{n+1}}{Z_n}.
\end{equation}
To estimate $Z$ using AIS, it is necessary to draw multiple unbiased samples from all $N-1$ intermediate distributions. Given that $N$ can be very large, a fast sampling method is essential.


\subsection{SimCIM numerical annealer}
\label{SIMCIMsec}
\begin{figure}[h] 
	\centering
	\includegraphics [width = 200 pt] {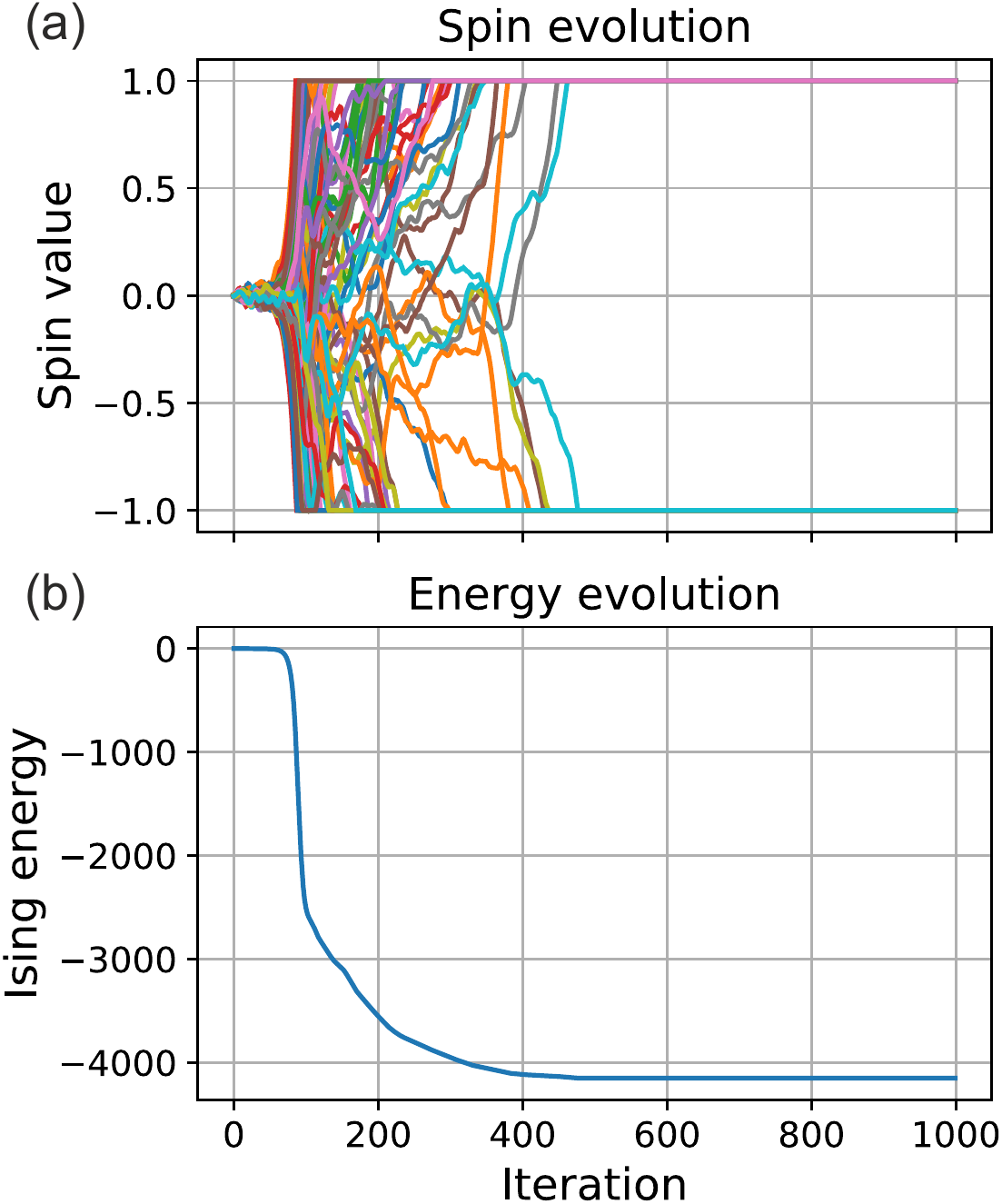}
	\caption{Application of a digital annealler to  the G6 graph from the G-set \cite{Gset}. a) Trajectories of individual spins. b) Evolution of the spin configuration energy for SimCIM.} 
	\label{NMFA_test}
\end{figure}

SimCIM is an iterative algorithm for sampling low-energy spin configurations in the classical Ising model. It treats each spin value $s_i$ as a continuous variable from the range $[-1, 1]$. Each iteration begins with calculating the mean field $\Phi_i = \sum_{j \neq i}J_{ij}s_j + b_i$ acting on each spin by all other spins. Then the gradients for the spin values are calculated according to $\Delta s_i = p_t s_i + \zeta \Phi_i + N(0,\sigma)$, where $p_t, \zeta$ are the annealing control parameters and  $N(0,\sigma)$ is Gaussian noise. Then the spin values are updated according to $s_i \leftarrow \phi(s_i + \Delta s_i)$, where $\phi(x)$ is the activation function
\begin{equation}\label{acti}
\phi(x)=\begin{cases}
x \textrm{ for } |x|\leq 1;\\
x/|x| \textrm{ otherwise}
\end{cases}
\end{equation}
After multiple updates, the spins will tend to either $-1$ or $+1$ and the final discrete spin configuration is obtained by taking the sign of each $s_i$. The typical evolution of spin values and the corresponding energy is shown in Fig.~\ref{NMFA_test}.
	
%
%
%

SimCIM has been tested on a variety of problems, including graphs from the G-set \cite{Gset}. Implemented on a consumer graphic processor, this algorithm runs faster and generates higher quality samples than many analog and digital annealing processes. 

\subsection{Routine for effective temperature estimation}
\label{Teffsec}

In 2016, Benedetti \textit{et al.} \cite{Benedetti2016} pointed out that quantum annealers that have strong interaction with the environment, such as D-Wave 2000Q, freeze out the dynamics of a spin system before the termination of the annealing process. As a result, such annealers sample from a Boltzmann distribution with some finite temperature. As we demonstrate in the next section, the samples generated by SimCIM have the same property. 

As we see from Eqs.~(\ref{energyBM}) and (\ref{probability}), scaling   the coupling matrix  $\hat{J}$ and bias vector $\textbf{b}$ is equivalent to scaling the  effective temperature $\beta$ by the same factor. This enables us  to control the temperature of the distribution from which the samples are drawn, which is necessary for many applications, including BM training. To take advantage of this capability, however, we also need a method that would allow $\beta$ to be measured. We describe this method below \cite{Benedetti2016}. 

 Consider a sample set corresponding to some inverse temperature $\beta_1$. The Boltzmann probability of a spin configuration with energy $E$ is $p_{\beta_1}(E) = g(E)\frac{e^{-\beta_1 E}}{Z_{\beta_1}}$, where $g(E)$ is the degeneracy of level $E$. For two energy levels $E$ and $E'$, separated by $\Delta E = E - E'$, the logarithmic ratio of the probabilities is given by
\begin{equation} \label{log_ratio}
l(\beta_1) = \mathrm{ln}\frac{p_{\beta_1}(E)}{p_{\beta_2}(E')} = \mathrm{ln}\frac{g(E)}{g(E')} - \beta \Delta E.
\end{equation}
Because of the unknown degeneracies, we cannot use the above equation to evaluate $\beta_1$ directly from a set of samples. However, one can obtain an additional set of samples with a different inverse temperature, $\beta_2$, by scaling the coupling matrix and bias vector by $\beta_1/\beta_2$. The difference of the log ratios for the two sets is then
\begin{equation} \label{delta_log}
\Delta l= l(\beta_1) - l(\beta_2) = \mathrm{ln}\frac{p_{\beta_2}(E') p_{\beta_1}(E)}{p_{\beta_2}(E) p_{\beta_1}(E')} = (\beta_2 - \beta_1) \Delta E.
\end{equation}
Now $\beta_1$ and $\beta_2$ can be estimated from the slope of the linear dependence between $\Delta l$ and $\Delta E$ and the  ratio $\beta_1/\beta_2$. 

For this method to work well, it is necessary that the two sets of samples have significant overlap, i.e. contain multiple samples within small energy intervals around both $E$ and $E'$. This can be achieved by judicious choice of $\beta_1/\beta_2$.

\section{Results}
\label{results_sec}
\subsection{Temperature evaluation for SimCIM}
\label{T_eval_sec}

The goal of this subsection is to demonstrate  SimCIM's capability to draw unbiased samples from a desired Boltzmann distribution, and that the methods of temperature evaluation and control described in section \ref{Teffsec} apply to SimCIM. We work with an Ising Hamiltonian with $\textbf{b} = \textbf{0}$ and $\hat{J}$ being a $16\times16$ symmetric matrix whose off-diagonal elements are uniformly sampled from $\left\lbrace -1,0,1 \right\rbrace $ and the diagonal elements are set to $J_{ii} = 0$. This relatively simple Ising problem allows exhaustive search over spin combinations, thereby enabling us to draw a sample set from the exact Boltzmann distribution, which can be used as a benchmark.
\begin{figure}[h] 
	\centering
	\includegraphics [width = 200 pt] {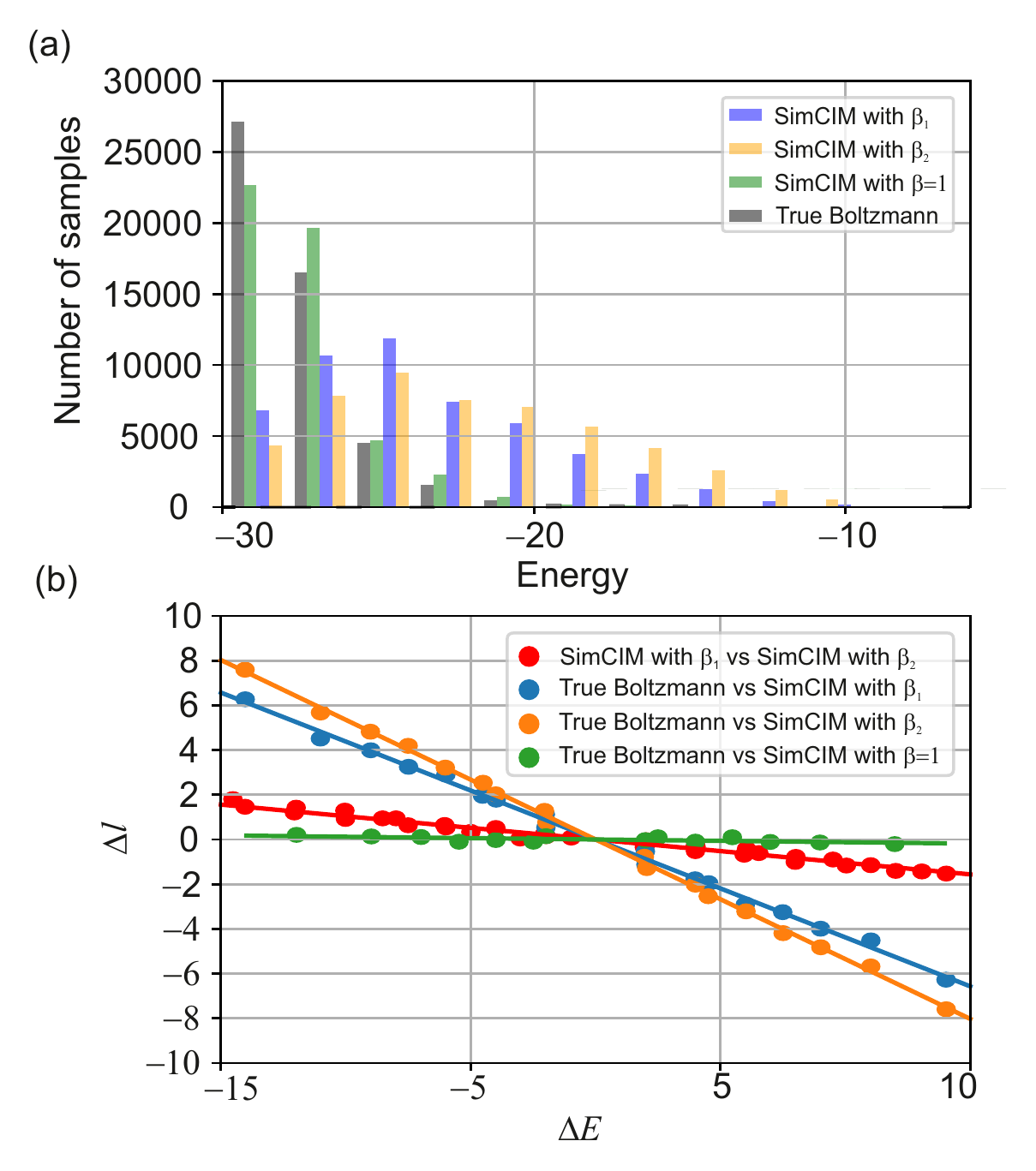}
	\caption{Temperature evaluation and control for SimCIM. (a) Histograms of 50000 vectors sampled either by SimCIM with different parameters $\beta$ or from the true Boltzmann distribution. (b) Dependencies \eqref{delta_log} for pairwise combinations of the sample sets from (a).} 
	\label{t_histograms}
\end{figure}

We sampled two batches of 50000 vectors via SimCIM with different inverse temperatures $\beta_1$ and $\beta_2$, with $\beta_1/\beta_2=0.76$, by scaling the coupling matrix $\hat{J}$. An additional set of 50000 samples was drawn from the true Boltzmann distribution with $\beta=1$. The histograms of these samples are presented in the Fig.~\ref{t_histograms} (a) and the pairwise dependencies $\Delta l$($\Delta E$), defined by Eq.~\eqref{delta_log}, in Fig.~\ref{t_histograms} (b). The linear shape of these dependences shows that the samples produced by SimCIM follow the Boltzmann distribution. 

Our next goal is to use SimCIM to sample from the Boltzmann distribution with $\beta=1$. To this end, we determine the slope of the dependencies $\Delta l$($\Delta E$) for the two sample sets drawn from SimCIM [red dots in Fig.~\ref{t_histograms} (b)] and find $\beta_1 - \beta_2=-8.75$, from which we find $\beta_1=28$ and $\beta_2=36.75$. Using this information, we rescale the coupling matrix for $\beta=1$ and sample 50000 additional vectors using SimCIM. The dependence  $\Delta l$($\Delta E$) for this set with respect to the true Boltzmann distribution is close to constant, as expected [Fig.~\ref{t_histograms}(b)]. 


\subsection{BM training on Bars \& Stipes dataset}
\label{BAS_BM_sec}

\begin{figure}[h] 
	\centering
	\includegraphics [width = \columnwidth] {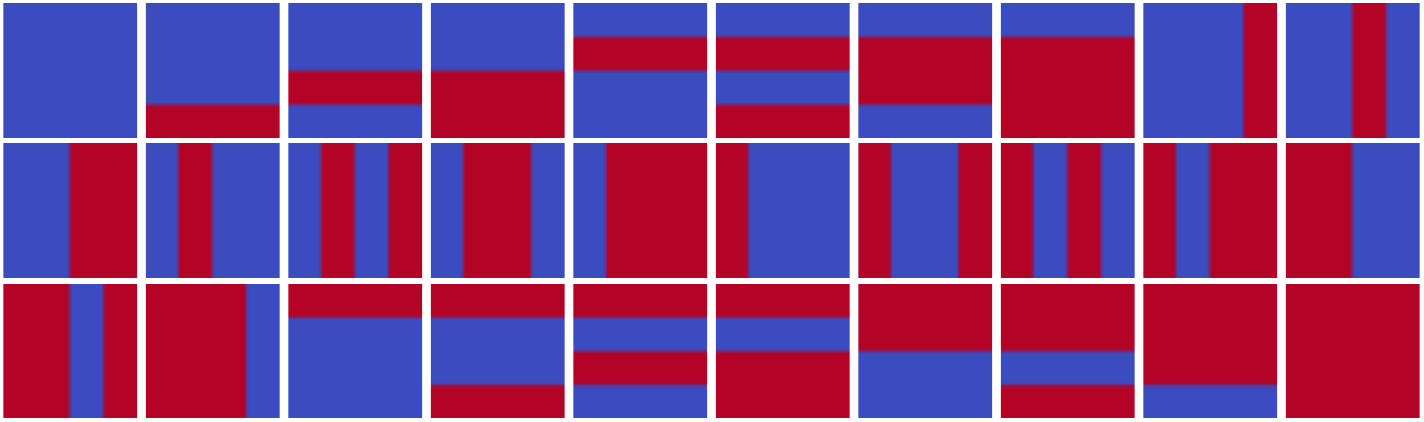}
	\caption{Bitmaps from the BAS dataset. Each picture is 4 $\times$ 4 pixels. Pixels are either -1 (blue) or +1 (red).} 
	\label{BAS_dataset}
\end{figure}

To benchmark the SimCIM performance as a sampling mechanism for machine learning, we start with BM training on the simple synthetic dataset from the Bars \& Stripes (BAS) family [Fig.~\ref{BAS_dataset}]. It consists of 30 instances of 4 $\times$ 4 bitmaps, with each pixel taking a value from $\left\lbrace -1, 1 \right\rbrace$. All instances occur with the same probabilities except the first and last one in Fig.~\ref{BAS_dataset}, whose probability of occurrence is twice as high. We refer to this as the ``ground truth probability distribution" of the training set. Similarly to the previous section, this simple dataset allows to do a full exhaustive search over  all spin configurations and is thus handy for benchmarking.  

The BM was trained using the method described in Sec.~\ref{BMsec}. Each bitmap was unrolled into a vector of size 16. The gradients' positive phase was always calculated from the full training set. The negative phase of gradients from Eqs.~(\ref{approx_updates}) was calculated using several different approaches. As a baseline method for comparison, we exactly calculated the gradients by exhaustive search of all possible spin configurations. The second method estimated gradients using samples drawn by MCMC with the chain length of 1000. Third, we utilized vectors sampled by SimCIM with and without temperature correction. All approximate methods of gradients' negative phase estimation used sample sets of equal size 250. To obtain temperature corrected sample sets, two additional sample sets of size 250 were produced for each update (see Sec. \ref{Teffsec}).

\begin{figure}[h] 
	\centering
	\includegraphics [width = 200 pt] {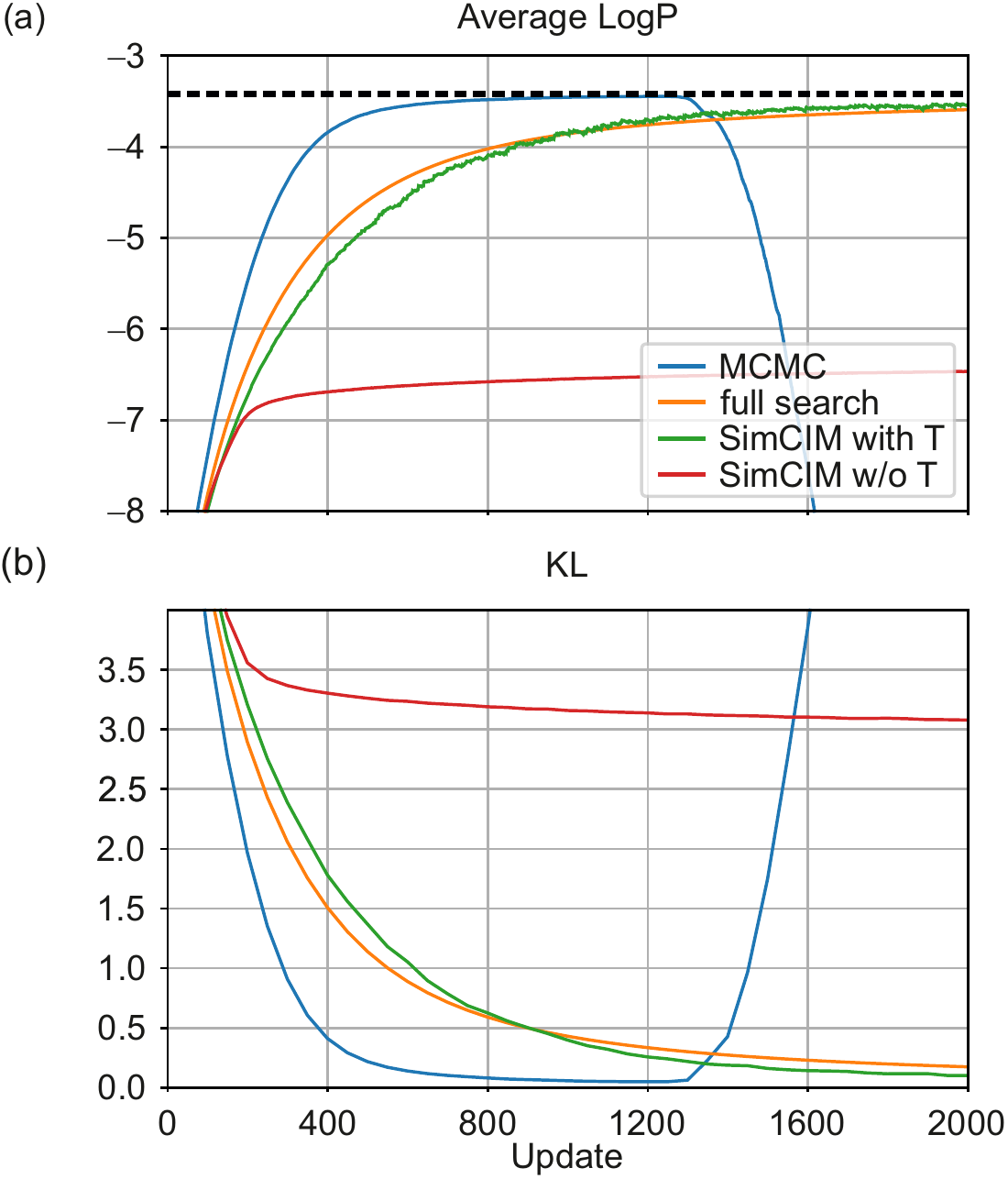}
	\caption{BM learning curves for the BAS dataset. (a) Average log-likelihood of the training set computed after each update of the BM parameters. The BM was trained by full exhaustive search (orange), vectors sampled by MCMC (blue), SimCIM with temperature compensation (green), and SimCIM without temperature compensation (red). The black dashed line corresponds to the ground truth average log-likelihood of the training set. (b) Kullback-Leibler divergence between the ground truth probability distribution defined by the training set and probability distribution defined by the current BM parameter set.} 
	\label{BAS_curves}
\end{figure}

To monitor the training process, we selected two numerical metrics. First, we calculated the mean log-likelihood of the training set with respect to the current state of the BM parameters. Second, we monitored the Kullback-Leibler (KL) divergence between the ground truth probability distribution defined by the training set and the distribution defined by the  current BM parameters, which can be calculated explicitly thanks to the small size of the problem. Both metrics were evaluated after each update of the BM parameters. The learning curves are presented in Fig.~\ref{BAS_curves}.

From these learning curves, we can see that gradients calculated using samples drawn by SimCIM with temperature correction follow the true gradients better than all other methods do. This is to be expected, because in order for the training method defined by Eqs.~\eqref{approx_updates} to work well, the samples must be drawn from the Boltzmann distribution with temperature 1. The vectors sampled by SimCIM satisfy this requirement better than samples provided by other sampling mechanisms.

\subsection{Partition function estimation}
\label{AIS_results_sec}


In this section, we implement AIS with SimCIM as the sampling tool to estimate the partition function $Z$ of a Boltzmann distribution and compare its performance with other approximate methods. The energy functionals for this test have been obtained from the coupling and weight matrices from consecutive epochs of the BM training process described in the previous subsection. The BM was trained using true gradient updates for 2000 epochs. The partition function was estimated every 50 epochs.

The methods included in the comparison are as follows. 
\begin{itemize}
    \item AIS equipped with SimCIM and MCMC samplers for importance weights calculations \eqref{AIS}. We used 10 intermediate distributions and sample sets of size 250 to compute each importance weight. Consecutive intermediate distributions differed from each other only by the inverse temperature $\beta$. To gradually transition from the uniform ($\beta = 0$) to target distribution ($\beta = 1$) we increase $\beta$ in $N=10$ steps. At each step, we used two additional sample batches of size 250 to estimate SimCIM's effective temperature. In total, AIS with SimCIM employed 7500 samples for the computation of each partition function, where only 2500 samples were used for the algorithm and rest are auxiliary samples for the temperature correction.
    \item VAN --- an artificial neural network-based method for partition function estimation of discrete distributions, which was sown to outperform mean-field family algorithms \cite{Wu2019}.  VAN was trained either for 2000 epochs with 250 samples in each batch (which gives $5\times10^5$ samples in total) or for 1500 epochs with batch-size of 5 (which is 7500 samples total). 
    \item Direct sampling, with the sample sets of size either 7500 or $5\times10^5$ drawn by SimCIM. 
\end{itemize}

\begin{figure}[h] 
	\centering
	\includegraphics [width = 200 pt] {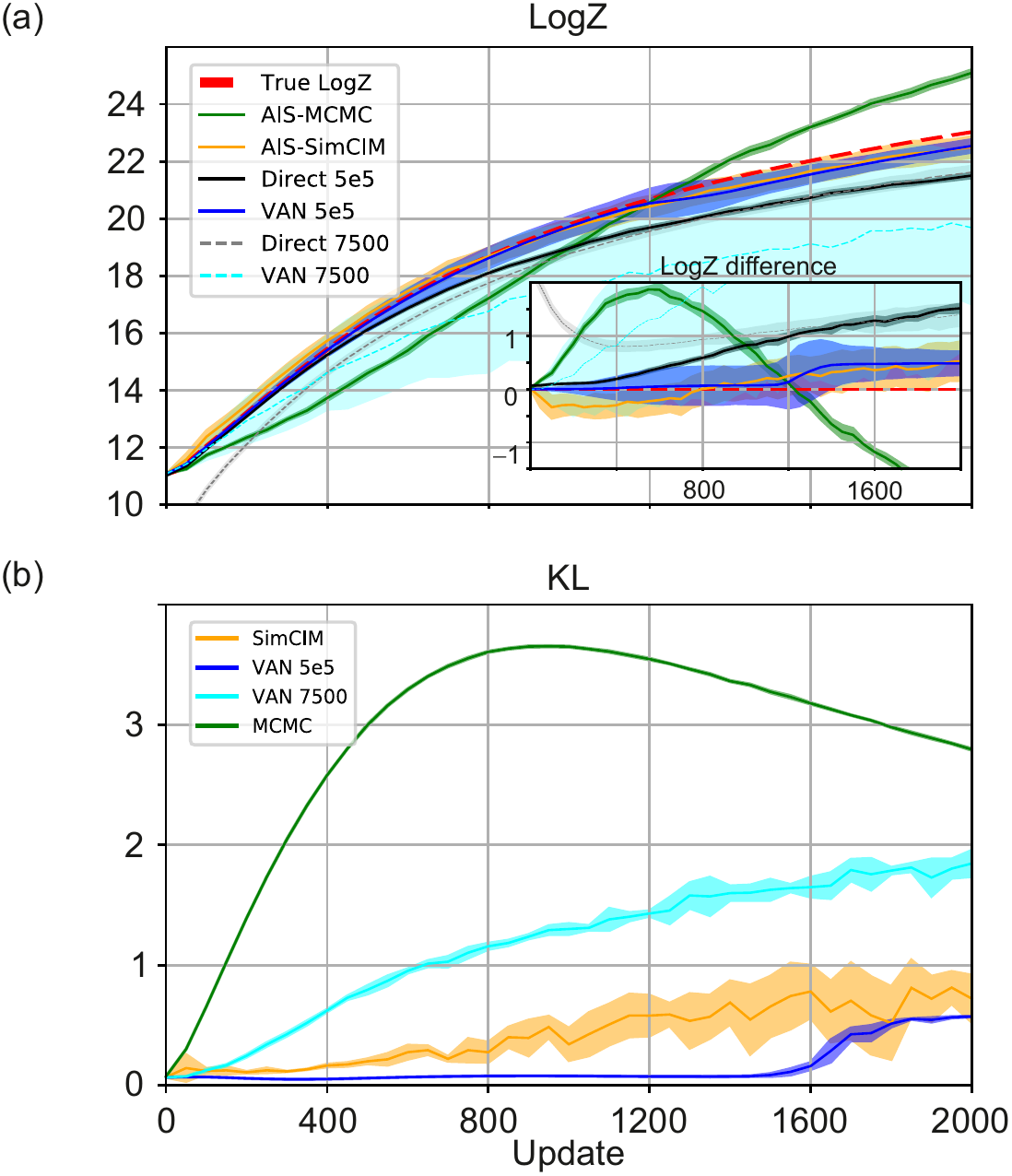}
	\caption{(a) Estimation of the logartithm $\log Z$ of the partition function LogZ for the probability distributions defined by the BM at different training epochs, implemented using several algorithms: AIS with MCMC (green) and SimCIM (orange) samplers; VAN with a total or 7500 (blue) and $5\times10^5$ (cyan, dashed) samples; direct sampling from SimCIM with 7500 (black) or $5\times10^5$ (gray, dashed) samples. The exact value (red) is shown for comparison. Inset: difference between the exact value and various approximations of $\log Z$. (b) KL divergence between the true Boltzmann distribution and its approximations derived from $5\times10^5$ vectors sampled by different algorithms.} 
	\label{AIS_VAN}
\end{figure}
The results are sown in  Fig.~\ref{AIS_VAN}. We see that the most competitive methods are AIS equipped with the SimCIM sampler and VAN; however, the total sample consumption with AIS+SimCIM was 7500 while that with VAN was $5\times 10^5$. 

Fig. \ref{AIS_VAN} (b) shows the KL divergence between the true Boltzmann probability distribution and approximate distributions derived using $5\times 10^5$ samples drawn by SimCIM, MCMC and trained VAN. We can see that VAN (which consumed more samples for training) approximates the true Boltzmann probability distribution better compared to all other methods. However, VAN requires training for every new BM parameter set, which could be quite time-consuming. For example, it takes around 15 seconds on a desktop with an NVidia GTX 1080 Ti GPU for VAN, while SimCIM samples 7500 vectors in 1.5 seconds. 

\begin{figure}[h] 
	\centering
	\includegraphics [width = 200 pt] {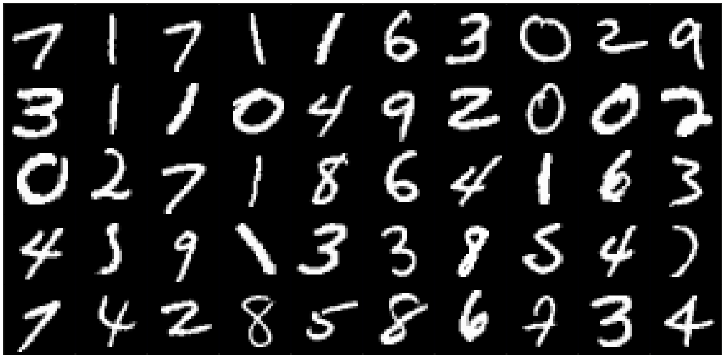}
	\caption{Bitmaps from the MNIST dataset.} 
	\label{MNIST_orig}
\end{figure}

\subsection{BM training on MNIST dataset}
\label{BM_results_sec}

To further test the capabilities of SimCIM as a sampling mechanism, we trained a BM on bitmaps from the MNIST dataset (Fig.~\ref{MNIST_orig}). This dataset contains 60000 instances of $ 28 \times 28 $ pixel grayscale bitmaps of handwritten digits in the training set and 10000 similar bitmaps in the test set. Each bitmap was unrolled into a vector and the vector elements after normalization were binarized to the values $\left\lbrace -1, 1 \right\rbrace $ with a threshold of 0.3. In order to enable the recognition of different digits, each training bitmap was augmented with a 10-bit one-hot vector $\textbf l_c$ such that  $l_i = -1$ for $i \neq c$ and $l_i = 1$ otherwise, where $c\in{0,\ldots,9}$ is the label of (i.e.~the actual digit corresponding to) the bitmap. In this way, we obtained a set consisting of 60000 binary vectors of length $28 \times 28+10=794$, which we split into training and cross-validation sets containing 50000 and 10000 images respectively.

During the training, we used minibatches of size 500 for the calculation of the positive phase of the gradient. The negative phase was calculated using vector sets of size 250 sampled by SimCIM with and without effective temperature adjustment. The model's parameters were updated after each minibatch. The BM initial weights were initialized as small random numbers. The training procedure was executed for 50 epochs, each epoch containing $50000/500=100$ updates. 

The training process has been evaluated using two benchmarks. 
The first one was the average log-likelihood of the test set bitmaps.
Because the exact value of this parameter is intractable to compute in this case, we utilized AIS equipped with the SimCIM sampler to approximate the partition function $Z$ and subsequently estimate the average log-likelihood.
The second metric was the classification accuracy of the test set digits. Each test bitmap was augmented with 10 variants of the one-hot vectors $\textbf l_c$ corresponding to  10 possible labels, and the corresponding 10 energy values $E(\mathbf{s}|c)$ were calculated. The label  $c^{\star} = \mathrm{argmin}_cE(\mathbf{s}|c)$ corresponding to the lowest energy yielded the model's prediction for the digit contained in the bitmap. The choice of energy, rather than log-likelihood, as the classification criterion, is justified in this case, because, for a given coupling matrix, the partition function is constant and hence the likelihood \eqref{probability} is a monotonic function of the energy.

\begin{figure}[h] 
	\centering
	\includegraphics [width = 200 pt] {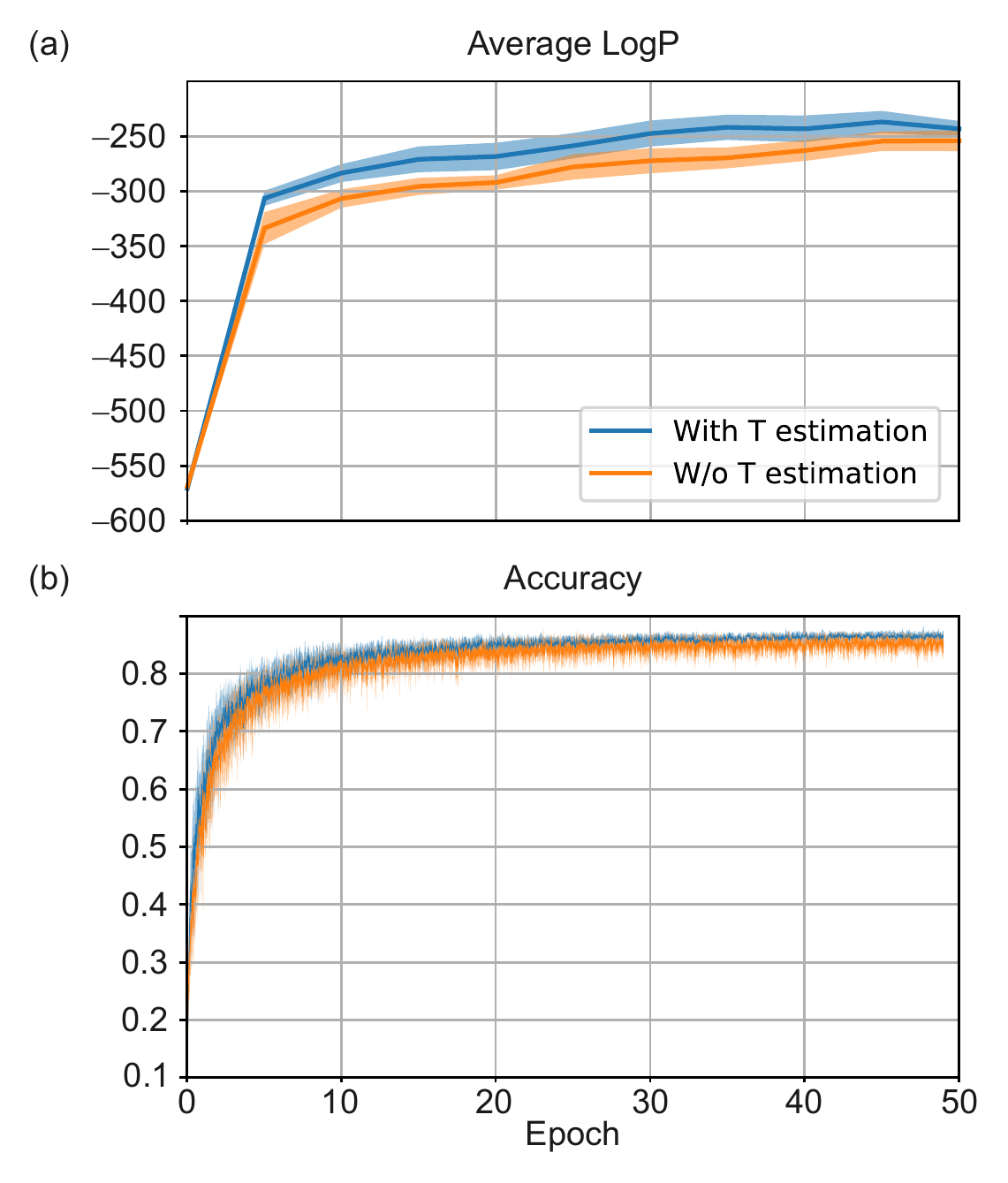}
	\caption{Learning curves for the approximate log-likelihood (a) and classification accuracy (b).} 
	\label{MNIST_acc_logp}
\end{figure}

The learning curves are shown in Fig.~\ref{MNIST_acc_logp}. We see that the BM can be trained successfully using the SimCIM sampler in view of both evaluation metrics. Consistently with our previous observations, the models trained using SimCIM with effective temperature adjustment perform better in terms of both the average log-likelihood and  classification accuracy. 

The classification accuracy reaches 86.9\% [Fig.~\ref{MNIST_acc_logp}(b)]. This result compares poorly to state-of-the-art supervised learning techniques based e.g.~on convolutional deep neural networks \cite{Ciresan2012} but is comparable to those obtained by other neural network architectures without hidden units. We believe that adding hidden units to our model would significantly enhance its accuracy.

In order to test the generative capabilities of the trained model, we used it to generate bitmaps corresponding to different digits. To that end, a one-hot vector $\mathbf{l}_c$ corresponding to a particular digit $c$ was chosen and the remaining part of the spin vector $\mathbf{s}$ was sampled using the coupling matrix obtained through the training. A fixed $\mathbf{l}_c$ gives rise to effective biases that ensure that bitmaps depicting the digit $c$ have lower energies, and are thus more likely to be produced by the annealer. 

\begin{figure*}[t!] 
	\centering
	\includegraphics [width = \textwidth] {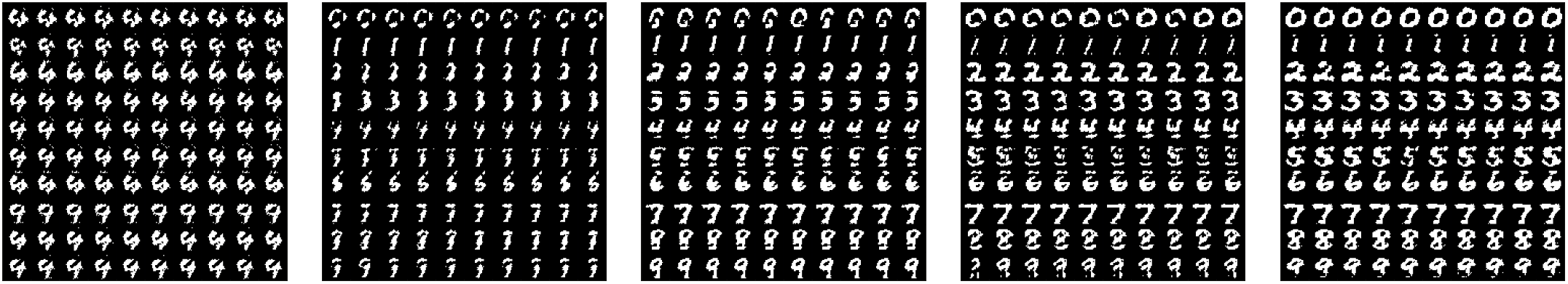}
	\caption{Digit reconstruction by the Boltzmann machine trained using the SimCIM sampler for (a) 4, (b) 9, (c) 14, (d) 29, and (e) 44 epochs respectively.} 
	\label{on_demand}
\end{figure*}

The results of this numerical experiment are shown in Fig.~\ref{on_demand}. We can see that the quality of generated digits improves with training.

\section{Summary}
We have demonstrated the quantum-inspired digital annealer SimCIM as a mechanism that provides high-quality one-shot samples from the Boltzmann distribution for a classical Ising energy functional. We used the samples to train a fully-visible and fully-connected general Boltzmann machine and to estimate the partition function of a high-dimensional probability distribution. The model trained on the MNIST dataset is able to achieve a reasonable classification accuracy and meaningful picture reconstruction. Approximate values of the partition function, as well as gradients computed using samples provided by SimCIM, agree well with the exact values computed using exhaustive search in a setting where such search is possible.

Implementation of SimCIM as a Boltzmann generator is beneficial compared to deep neural network-based methods because SimCIM does not require time-consuming training and could work in a plug-and-play manner. On the other hand, SimCIM is beneficial compared to quantum annealers such as D-wave because existing quantum annealing hardware suffers from multiple limitations, such as the noise in parameters, limited parameter range, restrictions in available architectures, decoherence, etc. \cite{Dumoulin2014}. Our algorithm enables simulation of various settings in which these limitations are curtailed or even completely eliminated, and therefore provides insights into the potential that near-term noisy intermediate-scale quantum (NISQ) devices \cite{Preskill2018} may have in application to various tasks such as ML or statistical physics.


This study motivates several open questions that could be of interest both for machine learning and quantum physics communities. ML researchers may wish to investigate whether or not quantum annealers and their simulators are applicable to the training of multilayer deep Boltzmann machines \cite{Salakhutdinov2009}, deep belief networks \cite{Hinton2006} or restricted BM based convolutonal neural networks \cite{Lee2011} and make rigorous comparison with existing methods of their training. For  physicists, the approach to BM training and partition function estimation presented here may be handy in the tasks of finding neural-network based solutions to quantum optimization problems, such as the ground states of molecules \cite{Xia2018} and quantum many-body systems \cite{Carleo2017} as well as quantum state tomography \cite{Torlai2018,TiunovQST2019} of high dimensional systems. A further interesting direction of future research is the development of analog annealers with the aim to outperform existing digital annealing algorithms. These annealers need not be limited to the quantum domain; indeed, analog annealers that successfully employ classical hardware have been proposed and tested \cite{Bello2019,Chou2019}.

\section*{Acknowledgments}
We are grateful to Jacob D. Biamonte for a fruitful discussion.
This work is supported by Russian Science Foundation (19-71-10092).

\end{document}